\documentclass[final,3p,times]{elsarticle}

\usepackage{amssymb}
\usepackage{amsmath}

\journal{Journal of Subatomic Particles and Cosmology}

\begin{document}

\begin{frontmatter}



\title{Light and heavy meson production in small collision systems}

\author[aaa]{Ivan Vitev}
\affiliation[aaa]{organization={Los Alamos National Laboratory, Theoretical Division},
             addressline={Mail Stop B283},
             city={Los Alamos},
             postcode={87545},
             state={NM},
             country={USA}}

\begin{abstract}
Recent results from the LHC on oxygen–oxygen (O-O) and neon–neon (Ne-Ne) collisions open a new window for investigating the interplay of cold 
nuclear matter (CNM) and quark–gluon plasma (QGP) effects in small collision systems. Building upon recent theoretical work on particle production 
dynamics in heavy-ion reactions, we present an updated study of light and heavy hadron modification relative to the proton-proton baseline in these 
systems for selected centralities. Our analysis combines perturbative QCD and hydrodynamic simulations to quantify initial-state effect, collisional energy 
loss, and medium-induced radiative corrections. We give theoretical predictions at both midrapidity and forward rapidity that can be confronted with 
ALICE, ATLAS, CMS, and LHCb measurements. Through comparison to the available data, we discuss the relative importance of CNM and QGP 
effects in O-O and Ne–Ne systems and the role of the heavy quark mass. Our analysis aims to clarify the onset of collective and deconfined behavior 
in small systems and to provide new insights into the transport properties of matter. We further argue that investigation of other observable such as 
energy correlators and quarkonia  can lead to a more complete picture of QGP formation in these collisions.
\end{abstract}


\end{frontmatter}

\section{Introduction}
\label{sec:intro}

Small collision systems have become central to the search for the smallest droplets of deconfined QCD matter that exhibits collective behavior~\cite{Ke:2022gkq,ATLAS:2016xpn,PHENIX:2023dxl,STAR:2026nfy,CMS:2025bta,CMS:2026qef}. In $p$--Pb, $d$--Au, and other asymmetric system reactions, long-range azimuthal correlations and flow-like patterns resemble signatures commonly associated with QGP formation in collisions of large nuclei. At the same time, the high-$p_T$ sector has not shown a comparably unambiguous jet-quenching signal. This tension motivates a systematic comparison of initial-state CNM effects and final-state QGP effects in systems whose size and geometry can be varied in a controlled way. O--O collisions are especially useful because they are symmetric, small, and experimentally accessible at RHIC and the LHC~\cite{STAR:2026nfy,CMS:2025bta,Strangmann:2026zxk}. They remove much of the uncertainty  of $p$--A centrality selection while still probing a medium whose lifetime and transverse extent are near the expected threshold for appreciable energy loss. Ne--Ne data extend this scan through a subtle change of the nuclear size while keeping the collision energy and detector environment similar~\cite{CMS:2026qef,LHCb:2026gwz}. 

Hadron spectra are inclusive probes of energy loss, whereas jet substructure observables can reveal how the lost energy is redistributed~\cite{Li:2017wwc,Ke:2025ibt,Ke:2023ixa,Lee:2022uwt}. The energy-energy correlator (EEC) measures the angular distribution of  particle-pair  energy flow inside a jet and is particularly sensitive to broadening and medium-induced radiation. We extend our study of small symmetric and asymmetric systems to in-medium EEC that include both fixed-order medium-modified splitting kernel contribution and renormalization group evolution  of this observable in nuclear matter~\cite{Ke:2025ibt,Ke:2023ixa,Ke:2024ytw}.

\section{Nuclear matter effects in small collision systems}
\label{sec:nucl}

{\em Inclusive hadron cross sections. } The analysis presented here uses a common baseline for light hadrons and open heavy flavor, following the high-$p_T$ hadron and heavy-flavor framework of Refs.~\cite{Ke:2022gkq,Ke:2023ixa}.  For the heavy-quark sector we use mass-dependent fragmentation and evolution, including explicit $m_q$ dependence of the in-medium splitting functions and the Lund--Bowler form for the nonperturbative fragmentation input. This is important because the finite masses of charm and beauty  change the branching pattern, reduce small-angle radiation, and make the relative importance of collisional and radiative energy loss more transparent. CNM contributions include a Cronin effect and we study the uncertainty associated with the presence or absence of cold nuclear matter energy loss. Final-state effects are accounted for through medium-modified DGLAP evolution, with couplings $g_{s}^{\rm med}=1.6$ and 1.8 motivated by the preferred range in prior light-hadron phenomenology, and through the inclusion of a collisional energy-loss component~\cite{Neufeld:2011yh}. Because collisional energy loss has a weaker path-length dependence than radiative energy loss in an expanding medium, it can become comparatively more important in small systems; this expectation is further tested through the mass hierarchy of $h^{\pm}$,  $D$, and $B$ meson suppression.

\begin{figure}[t]
\centering
\includegraphics[width=0.86\textwidth]{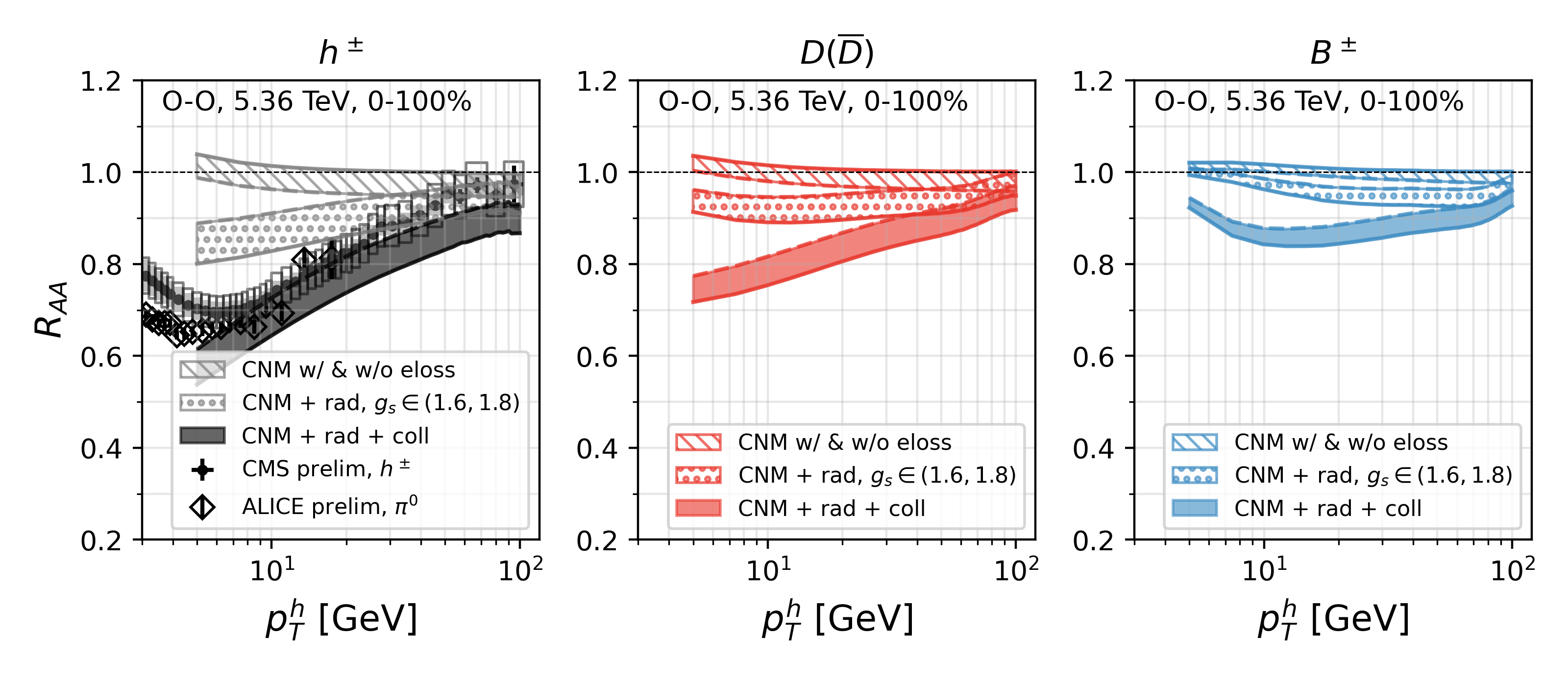}\vspace{-4.5mm}
\includegraphics[width=0.86\textwidth]{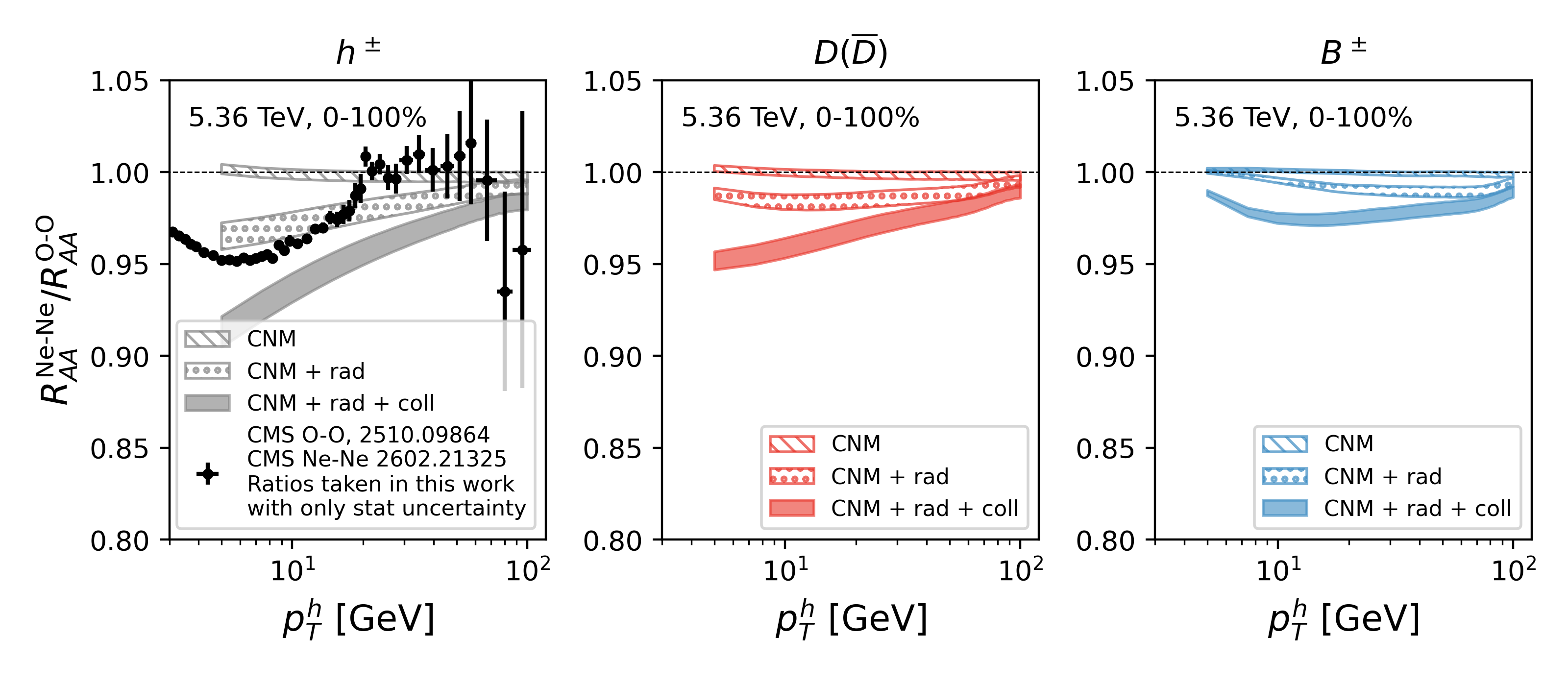}
\vspace{-3.5mm}
\caption{Top: nuclear modification factors in O--O collisions at $\sqrt{s_{NN}}=5.36$ TeV for charged hadrons, $D(\bar D)$ mesons, and $B^{\pm}$ mesons. Hatched bands show CNM effects with and without cold nuclear matter energy loss; dotted bands add medium-induced radiation with $g_s^{\rm med}=1.6$--1.8; filled bands also include collisional energy loss. Bottom: corresponding Ne--Ne/O--O double ratios. The charged-hadron panel includes CMS O--O data~\cite{CMS:2025bta}, ALICE $\pi^0$ data~\cite{Strangmann:2026zxk}, and ratios formed from CMS O--O and Ne--Ne preliminary measurements~\cite{CMS:2025bta,CMS:2026qef}. Our calculations follow Ref.~\cite{Ke:2022gkq}.}
\label{fig:hadrons}
\end{figure}

Figure~\ref{fig:hadrons} summarizes the resulting nuclear modification factors $R_{AA}$   for inclusive charged hadrons, prompt charm mesons, and beauty mesons in minimum-bias O--O collisions, together with the corresponding Ne--Ne/O--O double ratios. The charged-hadron and neutral-pion data shown in the figure are from CMS and ALICE O--O measurements~\cite{CMS:2025bta,Strangmann:2026zxk}; the double-ratio data are formed from the CMS O--O and Ne--Ne measurements~\cite{CMS:2025bta,CMS:2026qef}. The upper panels show that CNM effects alone produce comparatively modest modifications. We remark that this, except for a more pronounced Cronin enhancement,  was consistent with ATLAS $p$-Pb measurements~\cite{Ke:2022gkq,ATLAS:2016xpn}.  For small symmetric systems this clearly is not the case; charged-hadron and neutral-pion data lie in the region where additional final-state quenching is visible. Adding medium-induced radiation lowers $R_{AA}$, and the inclusion of collisional energy loss produces the strongest suppression at low and intermediate $p_T$.  Note, however, that strong collisional energy loss is not favored in this theoretical calculation. The lower panels of Fig.~\ref{fig:hadrons} show ratios of Ne--Ne to O--O nuclear modification factors. Because many experimental and theoretical normalization uncertainties cancel, these double ratios isolate the change of quenching with system size. The calculated ratios are near unity for pure CNM effects, while final-state effects lead to percent-level deviations whose sign and magnitude depend on flavor and momentum and are consistent with expectations. Charged-hadron data constructed from CMS O--O and Ne--Ne measurements present a ratio close to unity at large $p_T$, and also suggest smaller collisional energy losses. The heavy-flavor predictions indicate that charm and beauty ratios should be even closer to unity, providing a useful control for separating geometry-driven energy loss, and heavy quark mass from baseline CNM effects.  For forward rapidity in O--O, the hydrodynamic simulations account  for  $\sim$8\% lower charged-particle multiplicity relative to midrapidity. This small reduction does not noticeably change  the quenching of inclusive hadrons.

\begin{figure}[t]
\centering
\includegraphics[width=0.36\textwidth]{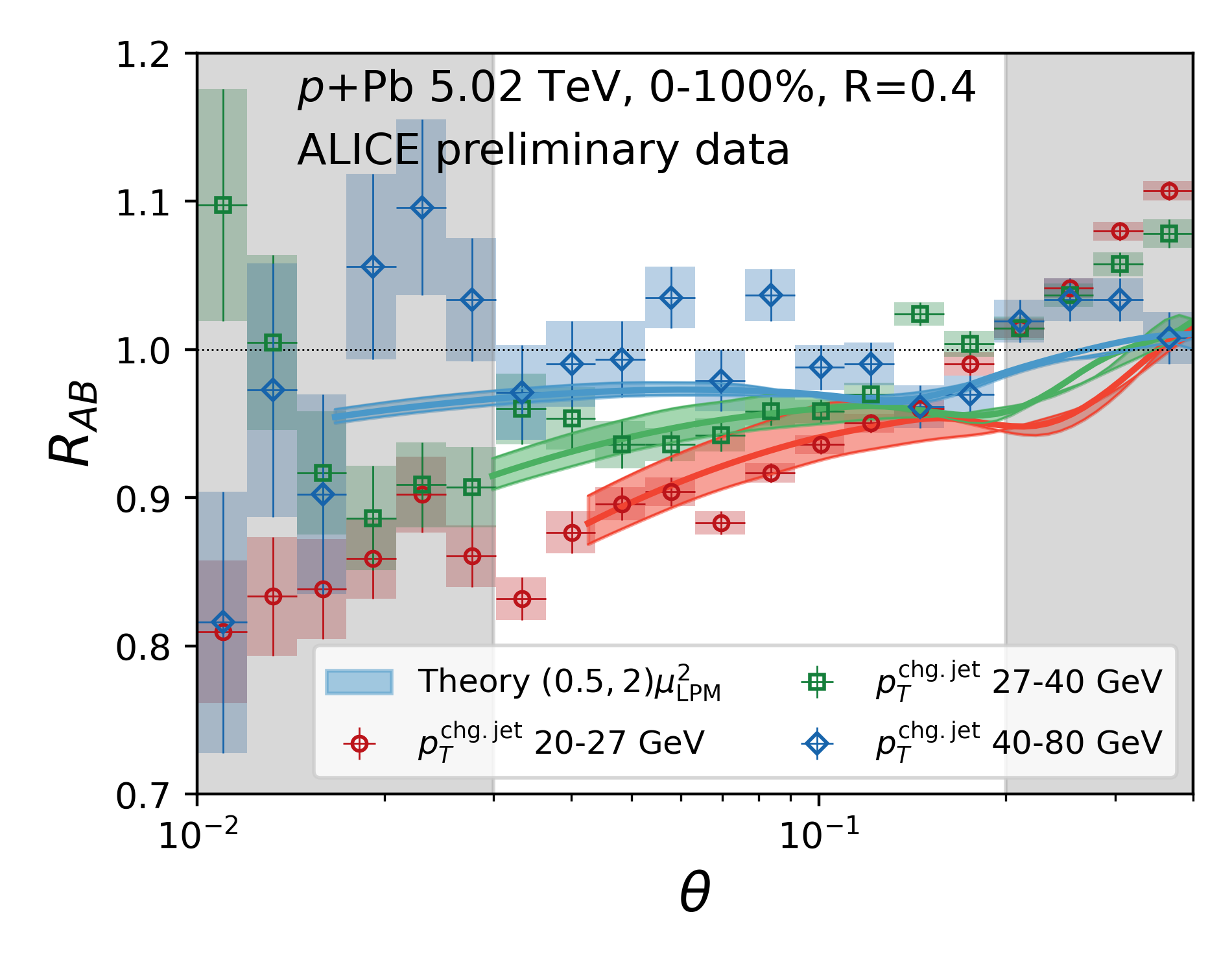}
\includegraphics[width=0.6\textwidth]{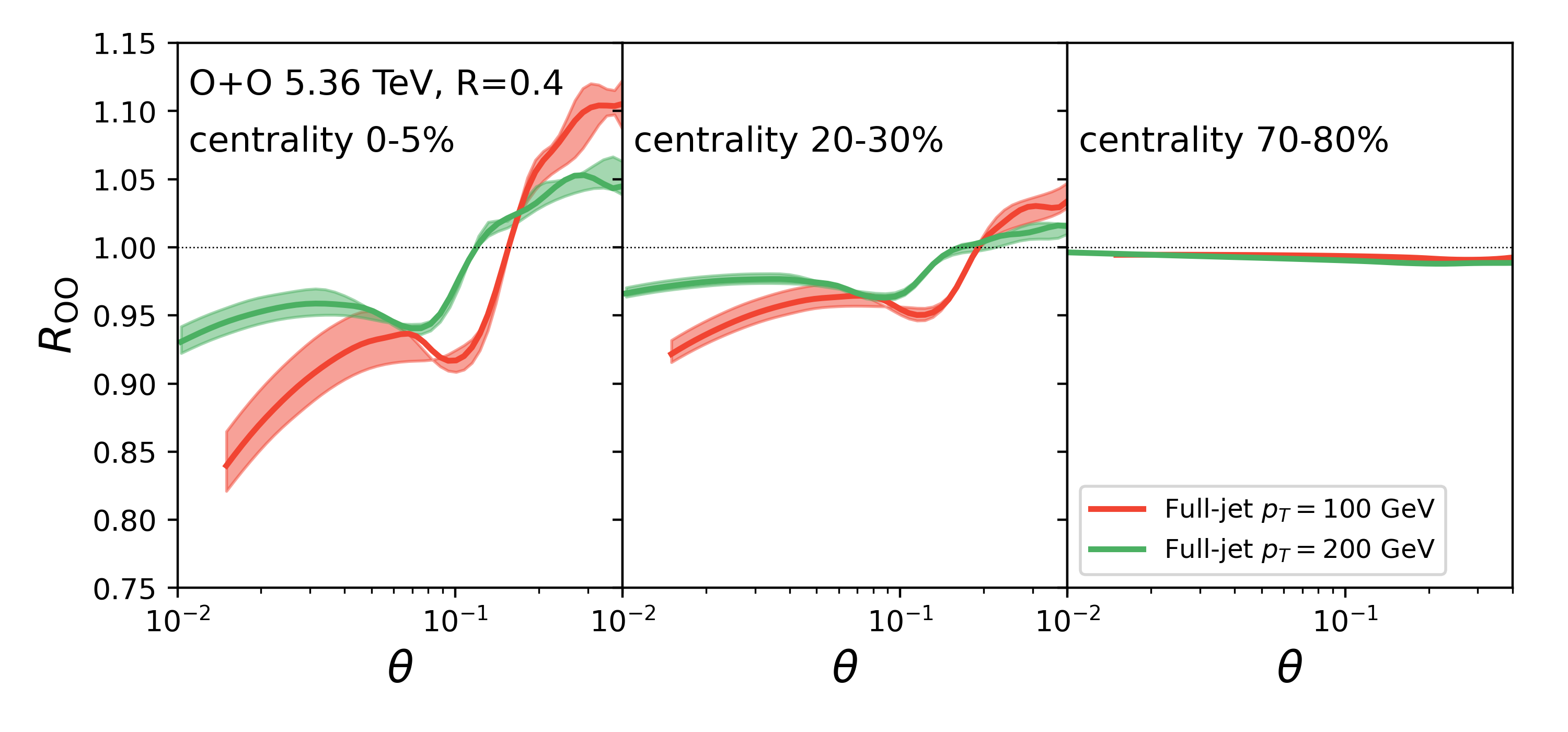}
\vspace{-3.5mm}
\caption{Left: nuclear modification factor $R_{pA}(\theta)$ of the EEC for charged jets in different $p_T^{\rm ch.\,jet}$ intervals in $p$--Pb collisions, compared with preliminary ALICE measurements~\cite{Liang-Gilman:2025gjl}. Bands denote hydrodynamics-averaged in-medium EEC calculations with first-opacity medium corrections and in-medium resummation~\cite{Ke:2025ibt,Ke:2023ixa}. Right: predicted O--O EEC modification $R_{\rm OO}(\theta)$ for full jets with $p_T=100$ and 200 GeV in central, midcentral, and peripheral events. Figures are from Ref.~\cite{Ke:2025ibt}.}
\label{fig:eec}
\end{figure}

{\em Energy-energy correlators. } Next, we turn to energy-energy correlators. In a recent work~\cite{Ke:2025ibt}, we incorporated nuclear medium interactions through a controlled opacity expansion that captures transverse momentum exchanges between hard partons and the QGP. Within this framework, we derived a factorized expression for the in-medium EEC and computed the quark and gluon jet functions that encode the leading effects of  hard scatterings on the collinear structure of the jet and provide a direct handle on medium-induced broadening and color decoherence. We further derived and analyzed the renormalization group (RG) evolution of the in-medium EEC, identifying how Glauber gluon exchanges modify the anomalous dimensions that govern the scale dependence of the observable. Our phenomenological results are given in  Fig.~\ref{fig:eec}: the left panel compares the EEC nuclear modification factor in $p$--Pb collisions with preliminary ALICE charged-jet data reported in Ref.~\cite{Liang-Gilman:2025gjl}. The calculation reproduces the qualitative pattern: suppression at small angles, a broad recovery around intermediate angles, and enhancement toward larger angles. This behavior  arises form the interplay of fixed order contributions  at moderate and large $\theta$ and evolution  due to out-of-cone radiation at small $\theta$. Shaded areas show regions where large uncertainties from non-perturbative effects such as fragmentation and  medium response may arise.  We note that the EEC modification in $p$-Pb is  consistent with the assumption of QGP formation,  while inclusive hadron quenching is not (but we emphasize again that the error bars are large).

The comparison also shows that measurements at higher jet momentum will be very valuable because the perturbative angular region is cleaner and less contaminated by hadronization ambiguities and underlying-event fluctuations. Furthermore, small symmetric systems will provide a more robust test of EEC modification in droplets of QGP. The right panel gives predictions for O--O collisions in three centrality classes and for two jet momenta, obtained with the in-medium framework of Refs.~\cite{Ke:2025ibt,Ke:2023ixa} and the same hydrodynamic medium treatment used for the hadron observables. As expected, the magnitude of the nuclear modification decreases with $p_T$ and centrality, but is sufficiently large in the more central events to pinpoint nuclear matter effects on parton shower formation.    


\section{Conclusions}
\label{Concl}

We presented an updated study of light- and heavy-flavor hadron production~\cite{Ke:2022gkq}, together with energy-energy correlators~\cite{Ke:2025ibt}, in small nuclear systems at collider energies. The results show that small symmetric systems such as O--O and Ne--Ne provide a particularly clean environment for separating cold nuclear matter effects from final-state interactions with a QGP droplet. In contrast to $p/d$--A reactions, where centrality determination and geometry biases remain significant, light-ion collisions preserve many of the advantages of a controlled nuclear-size scan while still probing matter close to the threshold for appreciable jet quenching.  The main conclusion of this study is that CNM effects and QGP effects lead to qualitatively and quantitatively distinct jet quenching patterns in small symmetric collision systems.  These differences are large enough to be tested by upcoming and preliminary measurements from ALICE, ATLAS, CMS, and LHCb. Heavy flavor is central to the interpretation of the underlying physics mechanisms. Charm and beauty production provide a mass-dependent test of parton shower modification and energy loss mechanisms. The predicted ordering between light hadrons, D mesons, and B mesons reflects the interplay of finite heavy-quark mass, medium-induced radiation, and collisional energy loss.

Similar to other jet substructure observables, EECs provide a complementary and more differential test of the underlying quantum chromodynamics (QCD) theory. The predicted suppression at small angular separation and enhancement at larger angle not only trace transverse momentum broadening and redistribution of energy within the jet, but also the RG evolution that arises from out-of-cone radiation. The centrality and jet-$p_T$ dependence in O--O, together with the comparison to $p$--Pb, can help determine whether the modification is controlled mainly by initial-state nuclear effects, by final-state interactions with a QGP droplet, or by a combination of both. Future extensions to higher logarithmic accuracy, higher orders in opacity, heavy-flavor jets, quarkonia, and multi-point correlators will strengthen the connection between jet substructure, heavy-quark transport, quarkonium production mechanisms, and the space-time evolution of QCD matter in the smallest systems accessible at collider energies~\cite{Copeland:2026yqa,Copeland:2025osx,Copeland:2025vop}.

\section*{Acknowledgments}
This researchs was supported by the LDRD program of Los Alamos National Laboratory. The author thanks Weiyao Ke and Bianka Me{\c{c}}aj for collaboration on the work presented here, and experimental colleagues for discussions of O--O, Ne--Ne, and EEC measurements.

\bibliographystyle{elsarticle-num}
\bibliography{sqm2026_template}



\end{document}